\documentclass[5p]{elsarticle} 

\usepackage{hyperref} %\usepackage{lineno,hyperref}
\usepackage{amssymb}
\usepackage{amsmath}
\usepackage{array}
\begin{document}

\begin{frontmatter}

\title{Image processing for grazing incidence fast atom diffraction.}

\author{Maxime Debiossac and Philippe Roncin}
\address{Institut des Sciences Mol\'{e}culaires d'Orsay (ISMO), CNRS, Univ. Paris-Sud, Universit\'{e} Paris-Saclay, F-91405 Orsay, France}


\begin{abstract}
Grazing incidence fast atom diffraction (GIFAD, or FAD) has developed as a surface sensitive technique. Compared with thermal energies helium diffraction (TEAS or HAS), GIFAD is less sensitive to thermal decoherence but also more demanding in terms of surface coherence, the mean distance between defects. Such high quality surfaces can be obtained from freshly cleaved crystals or in a molecular beam epitaxy (MBE) chamber where a GIFAD setup has been installed allowing in situ operation. Based on recent publications by Atkinson \textit{et al}\cite{Atkinson} and Debiossac \textit{at al}\cite{DebiossacPRB}, the paper describes in detail the basic steps needed to measure the relative intensities of the diffraction spots. Care is taken to outline the underlying physical assumptions.
\end{abstract}


\end{frontmatter}

%\linenumbers

\section{Introduction} 
MBE is the reference technique to produce very high quality crystalline surfaces while GIFAD is sensitive to surface coherence over length scales in the range of 1000$\AA$ \cite{DebiossacPRB,Busch2012,DebiossacPRL}. Recently, combining both techniques has increased the surface sensitivity of MBE diagnostics while providing surfaces with exceptional quality for \textit{in situ} fast atom diffraction. GIFAD can be operated easily both in growth condition where almost video rate was demonstrated \cite{Atkinson}, and in high resolution mode, before or after growth, where its sensitivity poses a challenge to the theoretical description \cite{DebiossacPRB}. The present paper focuses on the data analysis and the associated physics. To support the discussion the data presented here are taken from ref.\cite{DebiossacPRB} and \cite{Atkinson} correspond to a GaAs surface grown in situ by homo-epitaxy and held at a temperature of 570 $^\circ$C under an As$_4$ overpressure during the measurements.
\begin{figure}[ht]
	\begin{center}
		\includegraphics [width=70mm]{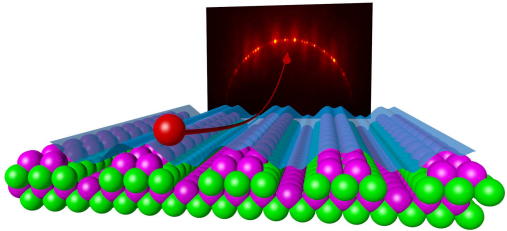}
	\end{center}
	\caption{{Schematic view of a GIFAD arrangement \cite{DebiossacPRB}, the primary beam of helium atoms does not really see individual atoms but are reflected by the periodic electronic density of  well-aligned rows of atoms.}}
	\label{schematic}
\end{figure}

\section{GIFAD and MBE}

\begin{figure}[ht]
	\begin{center}
		\includegraphics [width=80mm]{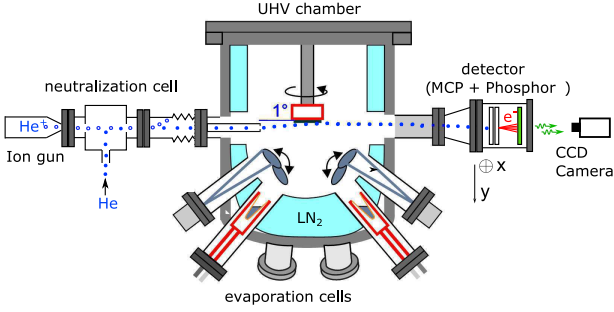}
	\end{center}
	\caption{{Schematic view of the MBE chamber \cite{Atkinson}, with effusion cells evaporating gallium and arsenic onto the GaAs(001) wafer. A He$^+$ ions beam is extracted  from a commercial ion source and is neutralized before entering the vessel. The atoms scattered by the surface are imaged onto a position sensitive detector.}}
	\label{MBE}
\end{figure}

GIFAD has been described in several places (see e.g. \cite{Winter_review}) and only a brief sketch is given here in fig.\ref{schematic} and \ref{MBE}. Just as HAS, helium atoms are weakly attracted by van der Waals forces and strongly repelled by the surface electronic density. In terms of interaction geometry, as seen on fig.\ref{MBE}, GIFAD is similar to reflection high energy electron diffraction (RHEED). In a way GIFAD is to HAS what RHEED is to low energy electron diffraction (LEED) as illustrated in Table \ref{LEED}. 

\begin{table} [h]%[<position>]
	\centering
    \begin{tabular}{|l|c|c|} 
	  \hline
	  geometry$\diagdown$ proj.  & electrons & He atoms \\
	  \hline
	  normal incidence& LEED & HAS,TEAS \\
	  grazing incidence& RHEED & GIFAD \\
	  \hline
    \end{tabular}
  \caption{{names of electron and atomic diffraction techniques}}
  \label{LEED}
\end{table}
An important practical difference is that keV atoms are used which can be detected with high efficiency and that the diffraction cone is kinetically compressed allowing the full pattern to fit onto a position sensitive detector. As a simplication, GIFAD can be seen as a projected technique where only the movement perpendicular to the surface plane is important \cite{Zugarramurdi,DebiossacPRA,Seifert_NIMB}. If the helium beam is well-aligned with this low index direction forming only an angle $\theta$ with the surface plane, the energy $E_\perp$ of the movement normal to the surface is given by $E_\perp=E_0 sin^2\theta$ with $E_0$ the energy of the primary beam. This energy $E_\perp$ can be tuned between few meV up to few eV with only a degree variation as illustrated in fig. \ref{raw_image}. Note that the wavelength $\lambda_\perp$ associated with this slow motion is in the \AA{} range. 

\section{Data analysis}
\subsection{primary beam, Laue circle and incidence plane}
\begin{figure}[ht]
	\begin{center}
		\includegraphics [width=80mm] {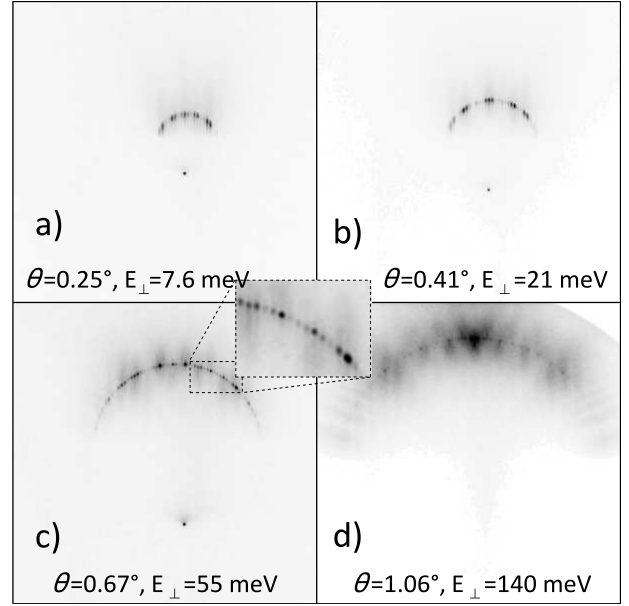}%{fig_raw.eps}
	\end{center}
	\caption{{Four diffraction patterns recorded for a $E_0$=400 eV He primary beam aligned along the [1-10] direction of the GaAs(001) surface held at 570 $^\circ$C corresponding to the $\beta_2$(2$\times$4) reconstruction \cite{DebiossacPRB}. The diffractions spots are located on the Laue circle which radius is equal to the angle of incidence $\theta$. From a) to d)  $\theta$ is increased from 0.25 deg. to 1.06 deg. allowing a factor close to 20 in normal energy  $E_\perp=E_0 sin^2\theta$ between the first and fourth image.}}
	\label{raw_image}
\end{figure}

We will not consider the nature of the position sensitive detector used to record the diffraction pattern as a 2D image. We assume here that the detector is far from the surface and perpendicular both to the surface plane and to the plane of incidence so that the number of counts in each pixel $(x,y)$ corresponds to an intensity map in momentum space $I(k_x,k_y)$. In the present case, one CCD pixel corresponds to a scattering angle of 0.004 deg. or 7 $10^{-5}$ rad. As with standard crystallography two type of information can be extracted, the surface lattice unit is reflected in the peak spacing while the relative peak intensities depend on the scattering amplitudes determined here by the shape of the electron density at the surface. The treatment starts by a precise determination of the primary beam parameters; its location x$_b$, y$_b$ and its width  which is measured here by its fwhm $\sigma_b=3.6$ pixel or $0.25$ mrad, symmetric along the x and y directions . This can be achieved before or after the diffraction by removing the target surface or, during diffraction by leaving a small part of the beam flying over the surface without interaction, as can be seen as a tiny spot at the bottom of figs.\ref{raw_image} a),b) and c). 
In favorable cases such as those depicted in fig.\ref{raw_image}, the Laue circle, defined by energy conservation ($|k_{out}|=|k_{in}|$) is clearly visible by interpolation between the elastic diffraction spots, and thus its center coordinates x$_L$, y$_L$ and radius R$_L$ are easy to pinpoint. However, this does not specify the scattering plane, defined on the detector by a line linking the primary beam to the specular beam\cite{DebiossacPRB,Lalmi}. If the direct beam is perfectly aligned with the low index crystal axis of the surface, the specular spot is easy to identify as the symmetry center but the alignment step can be tricky if no precautions have been taken \cite{Zugarramurdi,DebiossacPRA,Lalmi}. Here we assume that the scattering plane is identified as the vertical axis so that the horizontal axis of the image is parallel to the surface plane. The goal is now to extract the intensity along the Laue circle and to assign, as precisely as possible, the intensity of each diffraction spot.
\\
%$\sigma_b$=3.6 pixel fwhm et le fit mex  2.0 donne 3.5.(220 microns)

\subsection{Background subtraction}\label{Background§}
In fig.\ref{raw_image} and fig.\ref{Dy_image}a) some intensity is present below and above the Laue circle. Its contribution increases with the angle of incidence, i.e. with the normal energy $E_{\perp}$ from fig.\ref{raw_image}a) to fig.\ref{raw_image}d). This intensity, originating both from inelastic scattering \cite{Manson2008}, and from surface imperfections \cite{Pfandzelter,Michely}, is very interesting in itself providing information on the surface Debye temperature and on specific phonon properties. However, our concern here is only to derive the (periodic) topological properties of the surface by extracting the elastic diffraction signal, i.e. the intensity associated with the sharp spots sitting exactly on the Laue circle. As can be seen in fig.\ref{raw_image}, the background to be subtracted is far from uniform showing both diffraction features in the horizontal direction and pronounced patches along the vertical direction as well as a quasi-nodal structure along oblique directions \cite{Seifert_Young}.
\begin{figure}[ht]
	\begin{center}
		\includegraphics [width=70mm]{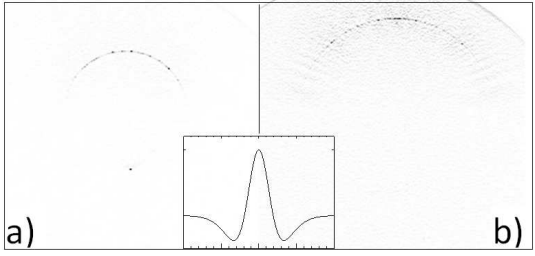}
	\end{center}
	\caption{{Same data as displayed in fig.\ref{raw_image}c) and \ref{raw_image}d) but filtered by a 1D mexican hat function (eq.\ref{eq_mex} and insert) in the vertical direction. The image looks empty but some of the diffraction spot are already saturated is gray scale.}}
	\label{image_mex}
\end{figure}
These structures originate from  quasi-elastic scattering processes \cite{Manson2008} which is closely linked to the elastic scattering \cite{Manson2008}. Similar patches are indeed present in diffraction charts when the diffracted intensities on the Laue circles \cite{DebiossacPRB,Schuller2008,Seifert_SiO2} are plotted as a function of the angle of incidence (see also fig.\ref{chart}). The reason is that the diffracted intensity can be written as an expansion in terms of the number of exchanged phonons. Close to the Laue circle, this number should be limited and the inelastic signal should resemble the elastic one in terms of intensity ratio of the various diffraction orders \cite{Gumhalter}. The fact that this background looks similar to the intensity on the Laue circle is a favorable condition for reasonable corrections. One way to estimate this contribution is to consider it as varying slowly in the vertical direction so that it can be interpolated from its values above and below the Laue circle by a linear interpolation. In practice, the background intensity $I_{back}(x,y)$ at any pixel $(x,y)$ on the Laue circle is estimated as $I_{back}(x,y)=[I(x,y+u)+I(x,y-u)]/2$ where the distance u will be chosen as small as possible but not smaller than the experimental resolution. This is equivalent to a double differentiation. To make this subtraction easy and not too noisy we use the smoothest possible filter; a 1D doubly differential "mexican hat" filter made of two gaussian functions (eq.\ref{eq_mex}) (see insert in fig.\ref{image_mex}) allowing a straightforward adjustment of the characteristic length u (eq.\ref{eq_mex}). 
\begin{equation}
	\ F_m(y)=2*e^{-0.5(y/\sigma)^2}-e^{-0.5(y/2\sigma)^2}
	\label{eq_mex}
\end{equation}
Here, a filter with a standard deviation $\sigma=1.4$ pixel in eq.\ref{eq_mex} is chosen giving a distance u of 3.3 pixels (0.23 mrad) between the center of the filter i.e. the positive pole and each of the negative poles. This value being close enough to the overall resolution taken here as the fwhm of the direct beam ; $\sigma_b$=3.6 channels = 0.25 mrad which is still significantly smaller than the fastest oscillation rate $\nu_{max}$. This frequency describes how fast a given diffraction order switches from bright to dark and then bright again during a rocking curve i.e. as the angle of incidence is varied (see again fig.\ref{chart}). For a beam energy of 400 eV on the $\beta_2$(2x4) reconstruction of the GaAs(001)\cite{Avery1996,LaBella1999} surface $\nu_{max}$ was measured at 0.9 \AA{}$^{-1}$ \cite{DebiossacPRB} corresponding here to approximately 1 mrad or a period P=15 CCD pixels on the detector. It means that hardly more than four significant data points (P/$\sigma$ 15/3.6) are available to describe this oscillation. It is therefore very important that the differential filter does not degrade the spatial resolution. Figure \ref{image_mex}a) shows that the application of this filter uniformly cancels the diffuse background isolating the intensity on the Laue circle. The effect of the filter is detailed further  in fig. \ref{Px}.

\subsection{Polar-like transformations}
\begin{figure}[h]
	\begin{center}
		\includegraphics [width=70mm]{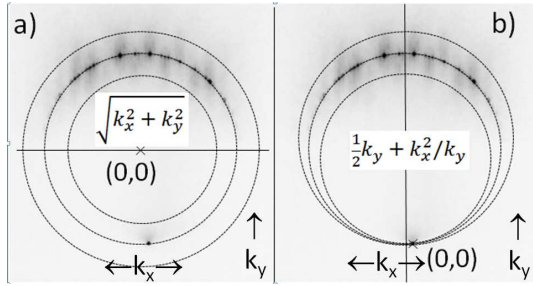}
	\end{center}
	\caption{{Schematic view of the two options for polar-like transformations. The reference is indicated by a cross corresponding to the origin. In a) the center of the Laue circle is the reference whereas in b) the reference is the direct beam. The equation for the respective circle radius $|\vec{k}_{out}|$ and  $k_{eff}$ are recalled in the insets.}}
	\label{circles}
\end{figure}

\begin{figure}[h]
	
	\begin{center}
		\includegraphics [width=60mm]{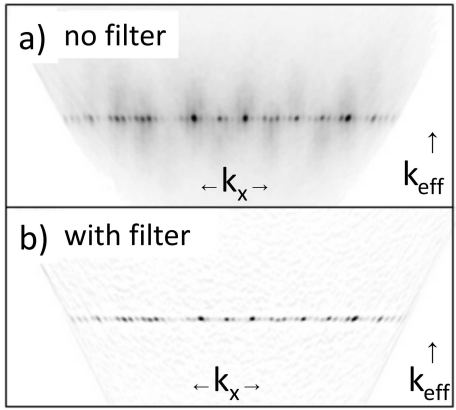}
	\end{center}
	\caption{{Same data as in fig\ref{raw_image}c), plotted now in ($k_x,k_{eff}$) coordinates i.e. after the polar like transformation. a) is without any filter showing a complex background structure while b) have been filtered in the vertical direction only by the same 1D filter as in fig.\ref{image_mex}.}}
	\label{Dy_image}
\end{figure}

We now need to convert the intensity on the Laue circle onto a line and this is exactly what a polar transform does.
However, a standard polar transform centered on the Laue circle and associating $(k_x,k_y) \rightarrow (\alpha,|\vec{k}_{out}|)$ with $\alpha =\arcsin k_x /k_y$ and $|\vec{k}_{out}|=(k_x^2+k_y^2)^{1/2}$ will not preserve the Bragg comb structure $k_x = n.G$.  Here n is the diffraction order, and $G=2\pi/a = 0.39 \AA^{-1} = 6.4 $ CCD pixels, the reciprocal lattice vector associated with the reconstructed lattice parameter $a=16\AA$. The obvious solution is to keep the $k_x$ coordinate intact and to consider the transformation $(k_x,k_y) \rightarrow (k_x,|\vec{k}_{out}|)$. To avoid a dual assignment, only the particles scattered above the surface plane i.e. only the values of $k_y$ above the center of the Laue circle $k_y>0$ are considered. For a well defined Laue circle as those reported in figs.\ref{raw_image} and \ref{image_mex}, this is enough to generate a 1D intensity distribution shown in fig.\ref{Px}. 
It should be noted that the Laue circle is not alway as easy to identify as in fig.\ref{raw_image}, this can be due to a limited coherence length of the surface defined here as the mean distance between crystallographic defects such as, ad-atoms, terraces, or to the excitation of surface phonons such as described by the Debye-Waller factor adapted to grazing incidence \cite{Manson2008,Rousseau2008}. This paragraph describes a more robust transformation ideally adapted to such common situations. 
It consists in using the primary beam location as a coordinate reference (0,0). In order to uniquely associate a circle to each pixel $(k_x, k_y)$ another information is needed. We use the specular scattering plane to generate a third symmetric point $(-k_x,k_y)$. The specular scattering plane is perpendicular to the surface plane and projects on the detector as a line linking the direct beam and the specular beam so that only one angle is needed. In practice this angle is the global image rotation angle that turns all diffraction images more or less symmetric with respect to the vertical axis. It has to be carefully recalculated for each new target surface or each new positioning of the camera but is not supposed to change otherwise. The circle associated with each point is now defined as the one encompassing itself $(k_x,k_y)$ , its mirror $(-k_x,k_y)$ and the direct beam (0,0). The radius of this circle defines the effective momentum $k_{eff}=k_y/2 + k_x^2/k_y$. The associated polar transform writes $(k_x,k_y) \rightarrow (k_x,2.k_{eff})$ as sketched in fig.\ref{circles}b) and illustrated in fig.\ref{Dy_image}. Note that the factor 2 is introduced so that the vertical line is invariant. For points located on the Laue circle, both approaches are strictly equivalent.  
Away from the Laue circle, the two transforms differ; when the Laue circle is taken as a reference (see fig.\ref{circles}a), the surface plane is implicitly taken as a perfect reference (all concentric circles stay in this plane) the angle of incidence is constant and perfectly defined ($|\vec{k}_{in}|$) but, away from the Laue circle outgoing scattering angle is different ($|\vec{k}_{out}|\neq |\vec{k}_{in}|$) which poses a serious problem if a normal energy and normal wavelength is to be defined for diffraction. At variance, in the second approach, the direct beam and scattered particle are located on the same circle so that an effective wave vector $k_{eff}$ or  angle of incidence can be defined as $\theta_{eff}=(\theta_{in}+\theta_{out})/2$ i.e. the circle radius in fig\ref{circles}b) allowing a quantitative analysis of the inelastic diffraction even out of the Laue circle.

The underlying assumption is that one has to consider the scattering on the surface as a quasi specular scattering taking place on a portion locally tilted by $\theta_{eff}-\theta_{spec}$ with respect with the macroscopic plane. This local tilt can be understood in classical terms as induced by the fact that an assembly of N atoms, each affected by thermal motion usually define a plane which is imperfectly aligned. This classical picture considers the surface composed of atoms frozen at their thermally displaced position during the scattering. In other words, any given ensemble of N atoms give rise to an effective local tilt that can be estimated by a linear regression between the given coordinates. 

In practice both models, while being strictly identical on the Laue circle, are also very similar as long as only the region around the Laue circle is concerned. Taking the beam as a reference is simply a more pragmatic choice with the additional benefit that for each circle an effective wavelength and wavenumber are associated which can be useful for quantitative analysis of inelastic diffraction. The result of this transform for fig.\ref{raw_image}c) is depicted in fig.\ref{Dy_image} where the vertical direction can now be interpreted as an effective wavenumber k$_{eff}$. The 1D vertical mexican hat filter described in section \ref{Background§} can be applied before or after the polar transform. Both are almost equivalent in the quasi specular region but give rise to a slightly different spot shape close to the equatorial plane, i.e. the intercept of the surface plane with the detector corresponding to the maximum exchange of momentum when the vertical momentum $k_{in}$ is entirely transformed into a horizontal momentum $k_{out}$ parallel to the surface plane.

\subsection{line profile}

\begin{figure}[ht]
	\begin{center}
		\includegraphics [width=80mm]{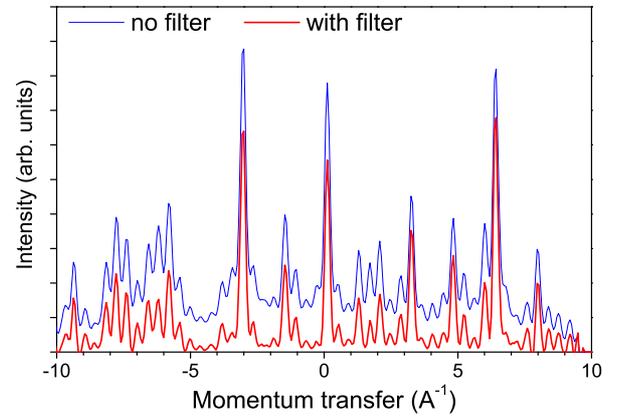}
	\end{center}
	\caption{{Intensity distribution along the Laue circle in fig.\ref{raw_image}c) corresponding to the central horizontal lines in fig.\ref{Dy_image}a) and b) respectively. When the filter is applied,the intensity is reduced but the contrast is clearly increased, note that the line shape is affected.}}
	\label{Px}
\end{figure} 
The intensity distribution on the Laue circle is displayed in fig.\ref{Px} with and without application of the mexican hat filter. As expected, the intensity is reduced but not uniformly. The interesting aspect is that the contrast is clearly improved ; some of the lines are almost extinguished while no negative intensity is generated (marginally at high k). A closer look also shows that the application of this vertical filter also affects the horizontal profile of the peaks which now have the same gaussian profile as the primary beam with exactly the same fwhm. Away from the Laue circle the peak profile is clearly Lorentzian. It should be noted that without subtraction of the background, the intensity profile on the Laue circle is poorly fitted by gaussians and much better by Lorentzians but the present analysis suggests that this is simply due to the presence of the background which generates significant intensity between the Bragg peaks.

\subsection{Diffraction chart}
\begin{figure}[ht]
	\begin{center}\includegraphics [width=80mm]{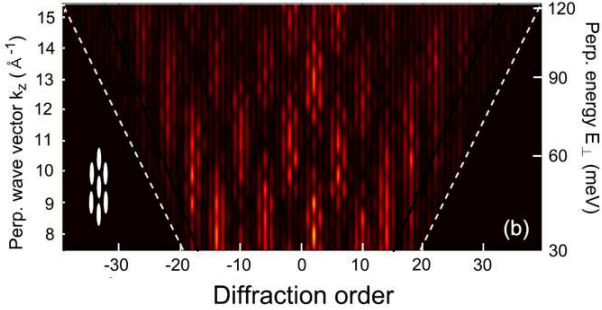}\end{center}
	\caption{{Diffraction chart taken from ref\cite{DebiossacPRB} constructed by juxtaposition of successive intensity distributions such as the one in fig.\ref{Px} at different angle of incidence. A  specific pattern is repeated several time and depicted in white on the left.}}
	\label{chart}
\end{figure}

The relative intensities of all the Bragg peaks contain all the information on the scattering of the projectile by the surface for a given incidence angle. This is enough to compare with detailed calculations but the sensitivity is so high that it is not easy to learn from the possible mismatch. A diffraction chart i.e. a plot summarizing all the intensity profiles associated with the angle of incidence have proven to be much more instructive\cite{DebiossacPRB}. This diffraction chart is simply a rocking curve where all diffraction orders are plotted. Figure \ref{chart} displays such a chart in which very distinct patterns can be identified. 
It has been shown\cite{DebiossacPRB} that this pattern can be explained qualitatively with only six straight line trajectories illuminating the top of the bumps or the bottom of the valleys of the potential energy surface. These produce specific path differences and the pattern becomes visible just by varying the wavelengths, i.e. the phase difference associated with these path differences. 

This simple model\cite{DebiossacPRB} is summarized here with a brief indication of the underlying assumptions.

-Firstly, the 3D potential energy surface $V(x,y,z)$ can be replaced by its 2D average $\tilde{V}(y,z)= \langle V(x,y,z) \rangle _x$ where x is the low index direction probed (here $[1\bar{1}0]$)\cite{Rousseau2007,Zugarramurdi}. Note that it can also be defined as the time average in the projectile frame $\tilde{V}(y,z)= \langle V(x,y,z) \rangle _t$

-Secondly, the well documented hard corrugated wall approximation considers that most of the momentum transfer occurs very close to the equipotential 1D function $\tilde{Z}(y)$ defined by $\tilde{V}(y,\tilde{Z}(y))= E_{\perp}$ so that straight line trajectories can be considered. As  in standard optics when a monochromatic light hits a diffraction grating, the diffraction can be calculated from a simple path difference between parallel rays. 

-The third simplification qualitatively considers that, in a restricted energy range, the shape of the equipotential function $\tilde{Z}(y)$ hardly varies with the perpendicular energy so that the diffraction chart can now be analyzed as the illumination of a fixed profile $\tilde{Z}(y)$ with a variable wavelength $\lambda_\perp$. This is valid when $E_{\perp} \gg E_{VdW}$ where $E_{VdW}$ is the typical Van der Waals energy. Note that this situation is almost never reached with thermal energy helium scattering but easily fulfilled with GIFAD.

\begin{figure}[ht]
	\begin{center}\includegraphics [width=70mm]{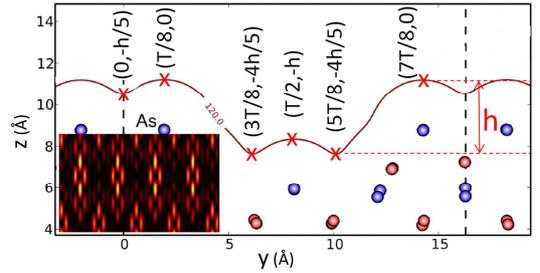}\end{center}
	\caption{{The equipotential line $\tilde{Z}(y)$ at 120 meV calculated by DFT ref\cite{DebiossacPRB} is replaced by the six point indicated by a cross  below their coordinates. T is the lattice unit and h=3.5 \AA{} is the full corrugation amplitude. inset: Diffraction chart produced by the ray tracing model with the six scattering centers.}}
	\label{pot}
\end{figure}

- Finally, in the quasi specular region, (particles bouncing almost vertically) the diffraction is considered to be dominated by the horizontal (flat) sections of $\tilde{Z}(y)$  i.e. the points $y_i$ such that $d\tilde{Z}(y_i)/dy =0$. These are the the top of the hills and the bottom of the valleys (fig.\ref{pot}) i.e. the natural points defining the atomic location.

The whole diffraction chart (inset in fig.\ref{pot}) can now be modeled with a reduced set of 6 coordinates ($y_j,z_j$), with $z_j = \tilde{Z}(y_j)$ inside the lattice cell and a straightforward analytical formula displayed in eq.\ref{rays} where the six corresponding, phase shifted, unit amplitudes associated with any diffraction order n are added and squared. 
\begin{equation}
\ I_n (k_{in})=|\varSigma_{j=1,6} \:e^{i\alpha_j (n)} |^2, \: \:\alpha_j (n)= nG*y_j + 2k_{in}*z_j. 	
\label{rays}
\end{equation}
Note that at $E_{\perp}$=120 meV the effective wavelength $\lambda_{\perp}$ is only 0.4$\AA{}$; almost ten times smaller that the corrugation amplitude h=3.5 $\AA{}$, the optical analog is therefore a grating with a very deep groove.

The interest of this approach is that it allows an analytical construction of a diffraction chart such as the one in inset of fig.\ref{pot} where the topological ingredients can be varied at will until a specific feature is identified. Such a simple approach is very handy to prepare more elaborate description of the surface, full confidence requires confirmation by a theoretical construction of the electronic density profile, and the use of exact diffraction codes such as wave-packet propagation\cite{DebiossacPRL,Rousseau2007,Aigner}, close-coupling \cite{DebiossacPRB,Zugarramurdi2015}, or even semi-classical approaches \cite{Gravielle}.

\section{GIFAD Images during growth}
One of the main applications of RHEED is the precise counting of the number of deposited layers. The two transformations presented above are simple enough to be applied with real time images, but we have focused more on the analysis after operation. The analysis of the GIFAD oscillations have been investigated in detail by Atkinson \textit{et al}\cite{Atkinson} with the homo-epitaxy of GaAs as a reference system. In this paper, it was shown that GIFAD oscillations are simple, robust and rich. Simple because the phase of the observed oscillations is the same for all diffraction orders, all primary energies, and angles of incidence. They indicate only the evolution of the surface reflectivity following the progression of a new layer. A maximum in a GIFAD oscillation indicates the completion of a new layer. The oscillations are robust for the same reason that, in layer by layer growth, the surface reflectivity oscillates even if diffraction features are not resolved. Finally, the GIFAD oscillations are rich because the helium scattering comprises three different components that were predicted \cite{Manson2008} and observed. These are the elastic diffraction, the inelastic or partly coherent diffraction, and the incoherent scattering.  
\begin{figure}[ht]
	\begin{center}\includegraphics [width=70mm]{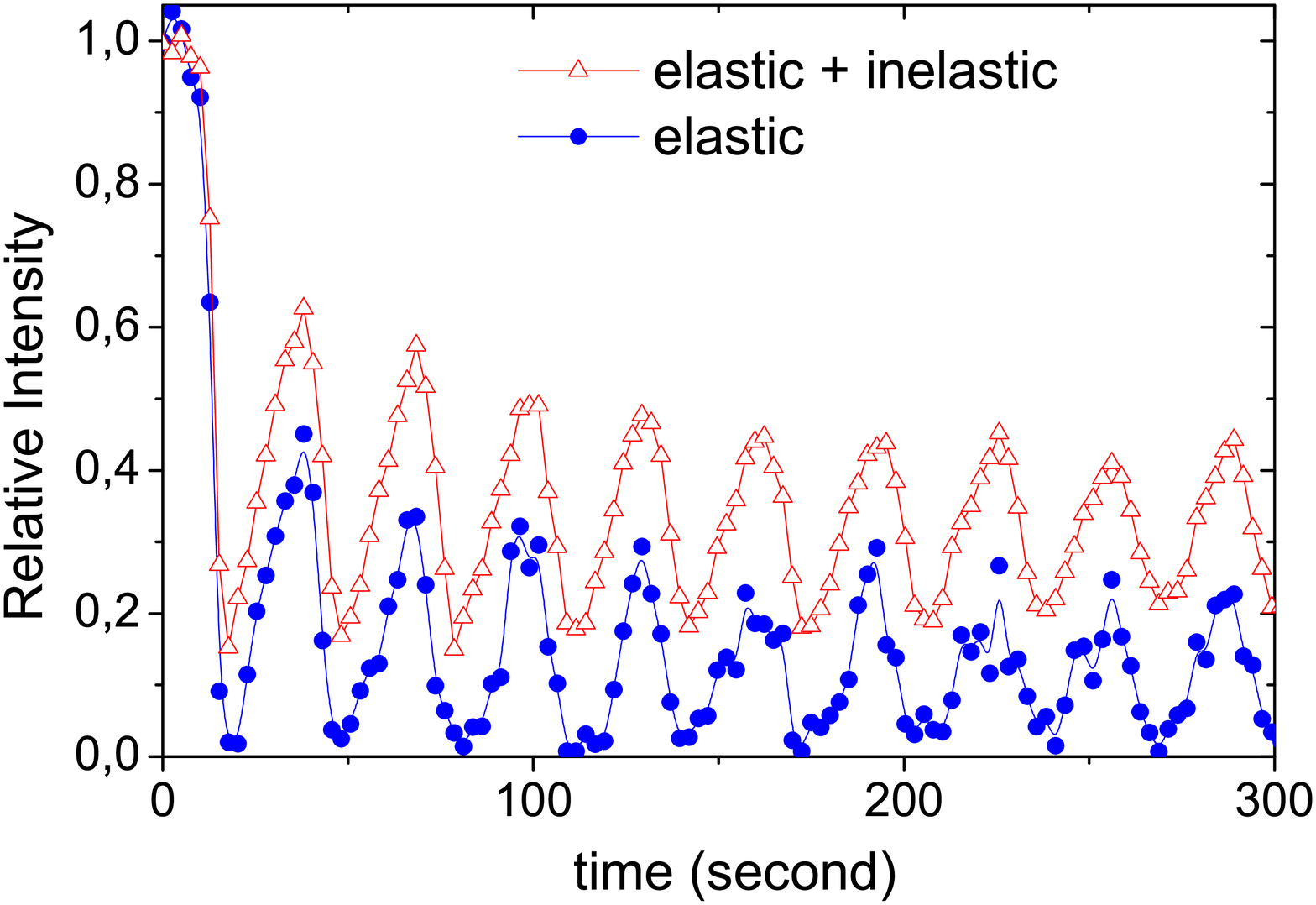}\end{center}
	\caption{{Time evolution of the intensity on the Laue circle during growth. It is referred to the intensity before growth. Full circles are for the elastic diffraction intensity while open symbol are for the total intensity including contributions from elastic, inelastic and incoherent scattering.}}
	\label{growth}
\end{figure}

The simple mexican hat filter described above was used to extract the elastic diffraction intensity and its evolution of during growth. Fig.\ref{growth} shows that the intensity of the elastic component drops to zero at the beginning of a new layer whereas the intensity on the Laue circle does not. Before growth, i.e. before opening the shutter in front of the Ga source the ratio of elastic diffraction, as measured with help of the mexican hat filter, is around 36\%. This is orders of magnitude larger than what is expected from the standard Debye-Waller factor for helium scattering from a GaAs surface at 570$^\circ$C. This reduced effective Debye-Waller is due to the multiple scattering centers involved in grazing angle scattering\cite{Rousseau2008,Manson2008}. The exact behavior of the three observed scattering (elastic, quasi-elastic and incoherent) intensities is still being investigated.
We propose here a tentative interpretation in terms of the sensitivity to  defects. Elastic diffraction probably requires that no defect is encountered during close contact with the surface while inelastic diffraction could survive as long as the momentum exchanged with defects is less than a reciprocal lattice vector, and incoherent scattering would pose no restriction but a is associated with a reduced probability to appear exactly on the Laue circle. More work is needed to better understand the exact length-scales associated with these contributions. As a result a model to fit the relative intensities should be possible which would provide data with less noise than the double differential filter used here. At variance, the advantage of the filter is that it is model independent. 

\section{Conclusion}
In order to isolate the intensity distribution on the Laue circle, two different polar-like transforms have been presented taking respectively the center of the Laue circle or the direct beam as references. Even in situation with a contrasted background is present above and below the Laue circle, a simple procedure is described to suppress the diffuse background on the Laue circle with a reduced statistical noise, allowing an improved contrast in the relative intensities. The diffraction spots have the same profile as the primary beam.

\section{Acknowledgement}
The data used here to illustrate the image processing were recorded during experiments funded by the Agence Nationale de la Recherche (Grants No. ANR-07-BLAN-0160-01). We are most grateful to P. Atkinson and M. Eddrief who provided the sample and operated the MBE chamber at the Institut des Nanosciences de Paris, while H. Khemliche and A. Momeni are kindly acknowledged for their help while running the GIFAD setup. We acknowledge the continuous and motivating theoretical support by A.G. Borisov A. Zugarramurdi and F. Finocchi.

%\lfoot{\small{\textsf{*Corresponding author e-mail address: philippe.roncin@u-psud.fr}}\\ \vspace{-0.5cm}\hrule}

\end{document}